\documentclass[a4paper,12pt]{article}
\usepackage{amsfonts,slashed}
\usepackage{url}
\usepackage{latexsym}
\usepackage{amsfonts}
\usepackage{epsfig}
\usepackage{latexsym,amssymb}
  \usepackage{amsmath,amssymb,amsthm}
\usepackage{ifpdf}
\ifx\pdfoutput\undefined
   \pdffalse
   \usepackage{cite}
 \else
   \pdfoutput=1
   \pdftrue
  \usepackage[pdftex]{hyperref}
\fi

\numberwithin{equation}{section}
\def\be{\begin{equation}}
\def\ee{\end{equation}}
\def\ba{\begin{array}}
\def\ea{\end{array}}

\newcommand{\bea}{\begin{eqnarray}}
\newcommand{\eea}{\end{eqnarray}}

\newcommand{\bbox}{\lower.2ex\hbox{$\Box$}}

\def\bfone{\relax{\rm 1\kern-.35em 1}}
\def\bfzero{\relax{\rm I\kern-.18em 0}}

\begin{document}
\begin{flushright}
CERN-TH-2016-051
\end{flushright}
\vskip 8mm
\begin{center}
{\bf\LARGE c-map for Born-Infeld theories} \\
\vskip 2 cm
{\bf \large L. Andrianopoli$^{1,2}$,  R. D'Auria$^{1}$, S. Ferrara$^{3,4,5}$
 and M. Trigiante$^{1,2}$}
\vskip 8mm
 \end{center}
\noindent {\small $^1$ DISAT, Politecnico di Torino, Corso Duca
    degli Abruzzi 24, I-10129 Turin\\
    $^{2}$  \it Istituto Nazionale di
    Fisica Nucleare (INFN) Sezione di Torino, Italy\\
    $^{3}$ Department of Theoretical Physics
CH - 1211 Geneva 23, SWITZERLAND\\
$^{4}$ INFN - Laboratori Nazionali di Frascati
Via Enrico Fermi 40, I-00044 Frascati, ITALY\\
$^{5}$ Department of Physics and Astronomy
U.C.L.A., Los Angeles CA 90095-1547, USA
}
\vskip 1.5 cm
\begin{center}
{\small {\bf Abstract}}
\end{center}
The c-map of four dimensional non-linear theories of electromagnetism is
considered both in the rigid case and in its coupling to gravity. In this way
theories with antisymmetric tensors and scalars are obtained, and the three non-linear representations of N=2 supersymmetry partially broken to N=1 related.
The manifest $\mathrm{Sp}(2n)$ and $\mathrm{U}(n)$ covariance of these theories in their multifield extensions
is also exhibited. This construction extends to H-invariant non-linear theories of Born-Infeld
type with non-dynamical scalars spanning a symmetric coset manifold G/H and the vector field strengths and their duals in a symplectic representation of G as is the case for extended supergravity.

\vskip 1 cm
\vfill
\noindent {\small{\it
    E-mail:  \\
{\tt laura.andrianopoli@polito.it}; \\
{\tt riccardo.dauria@polito.it}; \\
{\tt sergio.ferrara@cern.ch}; \\
{\tt mario.trigiante@polito.it}}}
   \eject
   \numberwithin{equation}{section}
\section{Introduction}
In this paper we formulate the c-map for Born-Infeld-like theories \cite{GZ} (for a review, see \cite{Aschieri:2008ns}), i.e. for non-linear theories  which generalize the canonical Born-Infeld (BI) electromagnetism to multi-vector, tensor and scalar fields. As for ordinary supersymmetric  theories, the c-map is defined \cite{Cecotti:1988qn}, \cite{Ferrara:1989ik}
both in the rigid and gravitational cases, by the uplifting of a formal dimensional  reduction of the four-dimensional theory on a circle to three dimensions \cite{{Breitenlohner:1987dg}}. Starting with a multi-vector generalization of the BI theory, the resulting three-dimensional model, obtained upon dimensional reduction and dualization of the vectors to scalar fields, is a non-linear model describing scalar fields only. This model provides a consistent non-linear theory for scalar fields in four dimensions, which in turn can be Legendre-transformed into a non-linear theory of antisymmetric rank-2 tensor fields. \par
This is most easily accomplished  using the linear description of BI theories, in which the Lagrangian is made quadratic in the vector field-strengths  by adding suitable Lagrange multipliers in the form of non-dynamical scalar fields \cite{Rocek:1997hi,Andrianopoli:2014mia}. Integrating  out these extra fields through their equations of motion, one obtains the BI action (or one of its multifield generalizations). One of the advantages of this formulation is that it makes the global symmetries of the BI theory manifest. Moreover it is the most convenient formulation in which to derive the BI-like theories from supersymmetric ones.

For the case of a single vector field, the three alternative formulations displayed in this paper, namely the scalar-scalar, scalar-tensor, tensor-tensor theories  \cite{Kuzenko:2000uh},\cite{Ferrara:2015ffa} are in fact related to different non-linear representations of the $N=2$ superalgebra spontaneously broken to $N=1$. While in the Born-Infeld case the goldstino multiplet is an $N=1$ vector multiplet \cite{Bagger:1996wp}, in the scalar-tensor theory it is an $N=1$ real linear multiplet \cite{Bagger:1997pi} and in the scalar-scalar case it is an $N=1$ chiral multiplet \cite{Bagger:1994vj}. The two latter theories correspond to D3 brane actions in five and six dimensions \cite{Bagger:1997pi}.

The paper is organized as follows. In section  \ref{BGM} we recall the basics of the c-map \cite{Cecotti:1988qn,Ferrara:1989ik} relating special and quaternionic (Hyper-K\"ahler) geometries in $N=2$ local  (rigid) theories. In section \ref{linear} we recall the main ingredients for the ``linear realization'' of BI-like theories, as developed in \cite{Andrianopoli:2014mia}.
In section \ref{cmapbi} we derive the c-map of the Born-Infeld theory coupled to gravity, which reproduces the bosonic part of the non-linear chiral multiplet action of \cite{Ferrara:2015ffa,Bagger:1997pi}. In section \ref{rigid} it is shown how all non-linear theories discussed in \cite{Bagger:1996wp,Bagger:1997pi} are reproduced using the c-map operation and Legendre transforms. In particular this implies that such theories have a  supersymmetric completion.
The paper ends with some concluding remarks.

\section{Local and rigid c-map}\label{BGM}
Let us recall, in this section, the formal steps to get  the c-map of the Lagrangian describing $N=2$ vector multiplets both in the supergravity and rigid-supersymmetry cases.
\paragraph{Local c-map.} Let us start from an $N=2$ supergravity model of vector multiplets in four dimensions \cite{Cecotti:1988qn} whose bosonic Lagrangian has the following general form:
\begin{equation}
 {e}^{-1}\mathcal{L}_4=-\frac{ {R}}{2}+g_{i\bar{\jmath}}\,\partial_\mu z^i\partial^\mu \bar{z}^{\bar{\jmath}}-\frac{1}{4}\, {F}^{\Lambda}_{ {\mu} {\nu}} g_{\Lambda\Sigma}\, {F}^{\Sigma\, {\mu} {\nu}}
+\frac{1}{4}\, {F}^{\Lambda}_{ {\mu} {\nu}} \theta_{\Lambda\Sigma}\,{}^* {F}^{\Sigma\, {\mu} {\nu}}\,,
\end{equation}
where $ {\mu,\nu}=0,1,2,3$, the index $\Lambda$ enumerates the vector fields and $ {F}^{\Lambda}_{ {\mu} {\nu}} =\partial_{ {\mu}} {A}^\Lambda_{ {\nu}}-\partial_{ {\nu}} {A}^\Lambda_{ {\mu}}$.
If the theory is invariant under axial rotations, we can formally perform a dimensional reduction, along the isometry direction, to three dimensions on a background with metric:
\begin{equation}
ds^2=e^{-2U}\,g_{\hat \mu\hat \nu}\,dx^{\hat \mu} dx^{\hat \nu}-e^{2U}\,(dx^3+A^{(3)})^2\,,
\end{equation}
where  ${\hat \mu},{\hat \nu}=0,1,2$ and $g_{{\hat \mu}{\hat \nu}}=g_{{\hat \mu}{\hat \nu}}(x^{\hat \rho})$, $A^{(3)}=A^{(3)}_{\hat \mu}(x^{\hat \nu}) dx^{\hat \mu}$ are the $D=3$ metric and Kaluza-Klein vector.
The vectors in $D=4$ reduce to three dimensional ones as follows:
\begin{equation}
A^\Lambda=\hat{A}^\Lambda_{\hat \mu}(x^{\hat \nu})\,dx^{\hat \mu}+\zeta^\Lambda(x^{\hat \nu})\,V^3\,\,,\,\,\,\,V^3=dx^3+\hat{A}^{(3)}\,.\label{43vec}
\end{equation}
where $V^3$ is proportional to the vielbein in the isometry direction and we have set $A^\Lambda_3= \zeta^\Lambda$.
The corresponding field strengths read:
\begin{equation}
F^\Lambda=\hat{F}^\Lambda+F^\Lambda_3\,V^3\,,
\quad \mbox{where }
\hat{F}^\Lambda=d\hat{A}^\Lambda+\zeta^\Lambda\,\hat{F}^{(3)}\,\,,\,\,\,\,\hat{F}^{(3)}=dA^{(3)}\,.
\end{equation}
Next we consider the $D=3$ Lagrangian which is given by the four-dimensional one written in terms of three dimensional fields, plus a Chern-Simons term inducing the dualization of the $D=3$ vector fields $A^{(3)}$, $\hat{A}^{\Lambda }$ to scalar degrees of freedom $a$, $\tilde\zeta_\Lambda$:
\begin{eqnarray}
{\hat e}^{-1}\mathcal{L}_3&=&-\frac{\hat{R}}{2}+\partial_{\hat \mu} U\partial^{\hat \mu} U-\frac{e^{4U}}{8} {\hat F}^{(3)}_{{\hat \mu}{\hat \nu}}\,{\hat F}^{(3)\,{\hat \mu}{\hat \nu}}+g_{i\bar{\jmath}}\,\partial_{\hat \mu} z^i\partial^{\hat \mu} \bar{z}^{\bar{\jmath}}+\nonumber\\
&&-\frac{e^{2U}}{4}\,{{\hat F}}^{\Lambda}_{{{\hat \mu}}{{\hat \nu}}} g_{\Lambda\Sigma}\,{{\hat F}}^{\Sigma\,{{\hat \mu}}{{\hat \nu}}}+\frac{e^{-2U}}{2}\,\partial_{\hat \mu} \zeta^{\Lambda} g_{\Lambda\Sigma}\,\partial^{\hat \mu} \zeta^{\Lambda}-\frac{1}{2\,e}\,\epsilon^{{\hat \mu}{\hat \nu}{\hat \rho}}\,{\hat F}^\Lambda_{{\hat \mu}{\hat \nu}}\theta_{\Lambda\Sigma}\partial_{\hat \rho} \zeta^\Sigma+\nonumber\\
&&+e^{-1}\mathcal{L}_{CS}\,,
\end{eqnarray}
where
$
\mathcal{L}_{CS}=\frac{1}{2}\,\epsilon^{{\hat \mu}{\hat \nu}{\hat \rho}}\,{\hat F}^\Lambda_{{\hat \mu}{\hat \nu}}\,\partial_{\hat \rho} \tilde{\zeta}_\Lambda-\frac{1}{4}\epsilon^{{\hat \mu}{\hat \nu}{\hat \rho}}\,{\hat F}^{(3)}_{{\hat \mu}{\hat \nu}}\,\omega_{\hat \rho}\,,
$
and we have defined $\epsilon_{{\hat \mu}{\hat \nu}{\hat \rho}}=\epsilon_{{\hat \mu}{\hat \nu}{\hat \rho} 3}$, so that $\epsilon_{012}=1$. The vector $\omega_{\hat \mu}$ is given in terms of scalar degrees of freedom and reads:
\begin{equation}
\omega_{\hat \mu}\equiv \partial_{\hat \mu} {a}+ \zeta^\Lambda\,\partial_{\hat \mu} \tilde{\zeta}_\Lambda-\partial_{\hat \mu} \zeta^\Lambda\,\tilde{\zeta}_\Lambda\,.
\end{equation}
Integrating out ${\hat F}^\Lambda_{{\hat \mu}{\hat \nu}}$ and ${\hat F}^{(3)}_{{\hat \mu}{\hat \nu}}$ we find the following equations:
\begin{eqnarray}
{\hat F}^{\Lambda\,{\hat \mu}{\hat \nu}}&=&-\frac{e^{-2U}}{e}\epsilon^{{\hat \mu}{\hat \nu}{\hat \rho}}\,g^{-1\,\Lambda\Sigma}(\theta_{\Sigma\Gamma}\,\partial_{\hat \rho} \zeta^\Gamma-\partial_{\hat \rho} \tilde{\zeta}_\Sigma)\nonumber\\
{\hat F}^{(3)\,{\hat \mu}{\hat \nu}}&=&-\frac{e^{-4U}}{e}\,\epsilon^{{\hat \mu}{\hat \nu}{\hat \rho}}\,\omega_{\hat \rho}\,.
\end{eqnarray}
Replacing the above solutions in $\mathcal{L}_3$ we find the final expression of the three dimensional Lagrangian fully written in terms of scalar degrees of freedom and exhibiting manifest $Sp(2n)$ structure \cite{Breitenlohner:1987dg}:
\begin{eqnarray}
{\hat e}^{-1}\mathcal{L}_3&=&-\frac{{{\hat R}}}{2}+\partial_{\hat \mu} U\partial^{\hat \mu} U+\frac{e^{-4U}}{4} \omega_{\hat \mu}\,\omega^{\hat \mu}+g_{i\bar{\jmath}}\,\partial_{\hat \mu} z^i\partial^{\hat \mu} \bar{z}^{\bar{\jmath}}+\frac{e^{-2U}}{2}\,\partial_{\hat \mu} \zeta^T g\,\partial^{\hat \mu} \zeta+\nonumber\\
&&+\frac{e^{-2U}}{2}\,\left(\partial_{\hat \mu} \tilde{\zeta}^T-\partial_{\hat \mu}\zeta^T\theta\right)g^{-1}\left(\partial^{\hat \mu} \tilde{\zeta}-\theta\partial^{\hat \mu}\zeta\right)=\nonumber\\
&=&
-\frac{{{\hat R}}}{2}+\partial_{\hat \mu} U\partial^{\hat \mu} U+\frac{e^{-4U}}{4} \omega_{\hat \mu}\,\omega^{\hat \mu}+g_{i\bar{\jmath}}\,\partial_{\hat \mu} z^i\partial^{\hat \mu} \bar{z}^{\bar{\jmath}}+\frac{e^{-2U}}{2}\,\partial_{\hat \mu}\mathcal{Z}^M\mathcal{M}_{MN}\partial^{\hat \mu}\mathcal{Z}^N\,,\label{localc}
\end{eqnarray}
being  $\mathcal{Z}^M=\{\zeta^\Lambda,\,\tilde{\zeta}_\Lambda\}$ and
\begin{equation}
\mathcal{M}\equiv \left(\begin{matrix}g+\theta g^{-1}\theta & -\theta g^{-1}\cr  -g^{-1}\theta& g^{-1}\end{matrix}\right)\,.\label{qual}
\end{equation}
Eq. (\ref{localc}) is the bosonic Lagrangian of $n+1$ hypermultiplets (containing the scalars $\{\zeta^\Lambda,\,\tilde{\zeta}_\Lambda$, $ z^i, \bar z^{\bar\jmath}, U, a\}$) coupled to gravity in $D=3$, $N=2$ supergravity, one of the multiplets (corresponding to $U, a$ and, say, $\zeta^0,\,\tilde{\zeta}_0$)   being the universal hypermultiplet containing the degrees of freedom of the supergravity multiplet in $D=4$. The scalars $(\mathcal{Z}^M,a)$ are acted on by the isometries
\begin{equation}
\delta\mathcal{Z}^M=\alpha^M\,,\quad\quad \delta a =\beta -\alpha^M \mathbb{C}_{MN} \mathcal{Z}^N\,,
\end{equation}
which close a characteristic Heisenberg algebra \cite{Ferrara:1989ik}.
 However, since hypermultiplets couple in the same way both to $D=3$ and to $D=4$ supergravity,  it can be promoted to a $D=4$ Lagrangian  describing the coupling of $n+1$ hypermultiplets to $D=4$ supergravity, by just extending the range of indices to $0,\dots,3$. The related scalar geometry is of quaternionic K\"ahler type \cite{Bagger:1983tt,deWit:1992wf}.

\paragraph{Rigid c-map.} This case is obtained from the local one by setting $g_{{\hat \mu}{\hat \nu}}=\eta_{{\hat \mu}{\hat \nu}}$, $U=\partial_{\hat \mu} U=0$ and $A^{(3)}={\hat F}^{(3)}=0$, that is $\omega_{\hat \mu}=0$.
Equation (\ref{43vec}), where now $\Lambda=1,\cdots ,n$, becomes:
\begin{equation}
A^\Lambda=\hat{A}^\Lambda_{\hat \mu}(x^{\hat \nu})\,dx^{\hat \mu}+\zeta^\Lambda(x^{\hat \nu})\,dx^3\,,
\end{equation}
and we define  ${\hat F}^\Lambda_{{\hat \mu}{\hat \nu}}=\partial_{\hat \mu} \hat{A}^\Lambda_{\hat \nu}-\partial_{\hat \nu} \hat{A}^\Lambda_{\hat \mu}$.
The $D=3$ Lagrangian reads:
\begin{eqnarray}
\mathcal{L}_3&=&g_{i\bar{\jmath}}\,\partial_{\hat \mu} z^i\partial^{\hat \mu} \bar{z}^{\bar{\jmath}}
-\frac{1}{4}\,{{\hat F}}^{\Lambda}_{{{\hat \mu}}{{\hat \nu}}} g_{\Lambda\Sigma}\,{{\hat F}}^{\Sigma\,{{\hat \mu}}{{\hat \nu}}}+\frac{1}{2}\,\partial_{\hat \mu} \zeta^{\Lambda} g_{\Lambda\Sigma}\,\partial^{\hat \mu} \zeta^{\Lambda}-\frac{1}{2}\,\epsilon^{{\hat \mu}{\hat \nu}{\hat \rho}}\,{\hat F}^\Lambda_{{\hat \mu}{\hat \nu}}\theta_{\Lambda\Sigma}\partial_{\hat \rho} \zeta^\Sigma+\mathcal{L}_{CS}\,,\nonumber\\&&
\end{eqnarray}
where
$
\mathcal{L}_{CS}=\frac{1}{2}\,\epsilon^{{\hat \mu}{\hat \nu}{\hat \rho}}\,{\hat F}^\Lambda_{{\hat \mu}{\hat \nu}}\,\partial_{\hat \rho} \tilde{\zeta}_\Lambda\,.
$
Solving with respect to ${\hat F}^\Lambda_{{\hat \mu}{\hat \nu}}$ we find:
\begin{eqnarray}
{\hat F}^{\Lambda\,{\hat \mu}{\hat \nu}}&=&-\epsilon^{{\hat \mu}{\hat \nu}{\hat \rho}}\,g^{-1\,\Lambda\Sigma}(\theta_{\Sigma\Gamma}\,\partial_{\hat \rho} \zeta^\Gamma-\partial_{\hat \rho} \tilde{\zeta}_\Sigma)\,.
\end{eqnarray}
Substituting in the Lagrangian we find:
\begin{eqnarray}
\mathcal{L}_3&=&g_{i\bar{\jmath}}\,\partial_{\hat \mu} z^i\partial^{\hat \mu} \bar{z}^{\bar{\jmath}}+\frac{1}{2}\,\partial_{\hat \mu}\mathcal{Z}^M\mathcal{M}_{MN}\partial^{\hat \mu}\mathcal{Z}^N\,.\label{hypl}
\end{eqnarray}
Analogously to the Lagrangian in the local case, given by eq. (\ref{localc}), eq. (\ref{hypl}) is the bosonic Lagrangian of $n $ hypermultiplets (corresponding to $\{\zeta^\Lambda,\,\tilde{\zeta}_\Lambda, z^i, \bar z^{\bar\jmath}\}$) in $D=3$, $N=2$ rigid supersymmetry, but it can be promoted to a $D=4$ Lagrangian  of the same form which describes an Hyper-K\"ahler sigma-model of a restricted type \cite{AlvarezGaume:1983ab,Cecotti:1988qn}.

\section{Linear description of Born-Infeld Theories}\label{linear}
In this section we briefly recall the linear  description of BI theories in terms of auxiliary fields, introduced in \cite{Rocek:1997hi,Aschieri:1999jr,Andrianopoli:2014mia}. This description does not rely on supersymmetry although, for special choices of the scalar sector and of the parameters, it can be embedded in a supersymmetric theory.
As extensively discussed in \cite{Andrianopoli:2014mia}, the  four-dimensional Lagrangian generalizing BI to $n$ vector fields can be put in the form (up to an additive constant):
\begin{equation}
\mathcal{L}=  -\frac{1}{4}\, {F}^T_{ {\mu} {\nu}}\,g\, {F}^{ {\mu} {\nu}}+
\frac{1}{4}\, {F}^T_{ {\mu} {\nu}}\,\theta\,{}^* {F}^{ {\mu} {\nu}}-\frac{1}{2\lambda}\,{\rm Tr}(N\mathcal{ M}) +const.\,,\label{linbi}
\end{equation}
where $N$ is a constant $2n\times 2n$ symmetric matrix,  $g$ and $\theta$ are $n\times n$ symmetric matrices function of a set of $n_s$ scalar fields $\phi^s$, $\lambda$ is a parameter which should be taken small to obtain a well-defined non-linear description.

The non-dynamical scalar sector can be integrated out through its algebraic equations of motion \cite{Andrianopoli:2014mia}, thus yielding a non-linear $n$-vector Lagrangian of BI type. These equations of motion can be cast in the following manifestly symplectic-covariant form:
\begin{eqnarray}
 \mathbb{F}^T_{\mu\nu} \partial_s \mathcal{M}\mathbb{F}^{\mu\nu}= -\frac 4\lambda
\mathrm{Tr}\left(N \partial_s \mathcal{M}\right)\,.
\end{eqnarray}
Here $\mathbb{F}=(F^\Lambda,\,G_\Lambda)$ is a symplectic vector built out of the electric  field strengths $F^\Lambda_{\mu\nu}$ and their magnetic duals $$G_{\Lambda\,\mu\nu}\equiv -\epsilon_{\mu\nu\rho\sigma}\frac{\delta\mathcal{L}}{\delta F^\Lambda_{\rho\sigma}}\,,$$ satisfying the field equations:
\begin{equation}
\partial_{[\mu}\mathbb{F}_{\nu\rho]}=0 \,; \quad {}^*\mathbb{F}_{\mu\nu}= - \mathbb{C}\mathcal{M} \mathbb{F}_{\mu\nu}\,,
\end{equation}
the latter being the so-called  ``twisted self-duality condition'' \cite{Cremmer:1979up}, and  $\mathbb{C}$ the $2n\times 2n$ symplectic invariant matrix $$\mathbb{C}=\left(\begin{matrix}{\bf 0} & {\bf 1}\cr -{\bf 1} & {\bf 0}\end{matrix}\right)\,.$$
The effective symmetry preserved by the non-linear Lagrangian depends on both the symmetry of the scalar sector and the invariance of the matrix $N$.
Suppose the scalar fields span a homogeneous symmetric space of the form $G/H$, and that the matrix $\mathcal{M}(\phi)$ defines a mapping between this manifold and $\mathrm{Sp}(2n)/\mathrm{U}(n)$:
\begin{equation}\{\phi^s\} \in \frac G H\,\,\,\longrightarrow\,\,\, \mathcal{M}(\phi)  \in \frac{\mathrm{Sp}(2n)}{\mathrm{U}(n)}\,.\label{mapghsp}\end{equation}
Any isometry generator of $G$, described by a Killing vector $k_\alpha$, corresponds to a symplectic matrix $(t_\alpha)_M{}^{N}$, so that
\begin{eqnarray}
\phi^s &\to& \phi^s + \delta \phi^s= \phi^s + \epsilon^\alpha k^s_\alpha \,:\,\,
\mathcal{M}\to \mathcal{M}+\delta \mathcal{M}\,,
\end{eqnarray}
with
\begin{equation}
\delta \mathcal{M}=\epsilon^\alpha k_\alpha^s   \partial_s \mathcal{M}=\epsilon^\alpha (t_\alpha \mathcal{M} + \mathcal{M} t^T_\alpha)\,.
\end{equation}
The on-shell global invariance of the  non-linear theory is described by the generators $t_\alpha$  of $G$ further satisfying the following conditions:
\begin{equation}
k_\alpha^s \mathbb{F}^T_{\mu\nu} \partial_s \mathcal{M}\mathbb{F}^{\mu\nu}= 2 \,  \mathbb{F}^T_{\mu\nu}  t_\alpha \mathcal{M}\mathbb{F}^{\mu\nu}=0 \,; \quad t_\alpha N + N t_\alpha^T=0\,.\label{eqf}
\end{equation}
These conditions define the group $G\bigcap\mathrm{Inv}(N)$ \cite{Andrianopoli:2014mia}, where $\mathrm{Inv}(N)\subset \mathrm{Sp}(2n)$ is the invariance group of the metric $N$.
In the case $N^{MN}=\delta^{MN}$, which is the choice we will make in what follows, $\mathrm{Inv}(N)=\mathrm{U}(n)$ and the global symmetry of the non-linear theory is the maximal compact subgroup $H$ of $G$.
Using the twisted self-duality condition, the first of (\ref{eqf}) can be cast in the form:
\begin{equation}
\mathbb{F}^T_{\mu\nu} t_\alpha \mathbb{C} {}^*\mathbb{F}^{\mu\nu}=0\,.
\end{equation}
These reproduce, in a symplectic invariant way, the conditions first found in \cite{GZ}.\par
Using the above setting, we can associate with each extended supergravity model, with $n$ vector fields and a symmetric scalar manifold $G/H$, a non-linear Born-Infeld theory featuring an on-shell symmetry $H$. This is done by adding to the bosonic Lagrangian the $H$-invariant potential $\frac{1}{2\lambda}\,{\rm Tr}(\mathcal{M})$ and dropping the kinetic terms of the scalar fields, so that they become non-dynamical. The map (\ref{mapghsp}) is built-in the mathematical structure of extended supergravities and is defined by the embedding of $G$ inside ${\rm Sp}(2n)$ \cite{GZ}. The symplectic matrix $\mathcal{M}$ has the general form $\mathcal{M}=LL^T$, where $L$ is the ${\rm Sp}(2n)$-representation of the coset representative. The non-linear BI theory originates by integrating the scalar fields  out through their algebraic equations of motion.\par
In the following we shall consider the case $G={\rm Sp}(2n)$, $H={\rm U}(n)$ and $N^{MN}=\delta^{MN}$. We postpone to a future work the study of non-linear theories with a smaller on-shell symmetry group, obtained by
considering the non-dynamical scalar fields in a smaller coset $G/H$.

\section{c-map of BI+gravity} \label{cmapbi}
We start from the linearized form of 1-vector BI coupled to four dimensional gravity which is obtained by coupling, for $n=1$, the Lagrangian (\ref{linbi}) to gravity:
\begin{equation}
\mathcal{L}= {e}\left(-\frac{ {R}}{2}-\frac{1}{4}\, {F}_{ {\mu} {\nu}}\,g\, {F}^{ {\mu} {\nu}}+
\frac{1}{4}\, {F}_{ {\mu} {\nu}}\,\theta\,{}^* {F}^{ {\mu} {\nu}}-\frac{1}{2\lambda}\,{\rm Tr}(\mathcal{M})+\frac{1}{\lambda}\right)\,,
\end{equation}
where  $ {\mu}, {\nu}=0,1,2,3$ and $\mathcal{M}$ was defined in Eq. (\ref{qual}).\par
Upon dimensional reduction on a circle and dualization of vectors to scalars, as discussed in Section \ref{BGM}, we end up with  a 3D hypermultiplet Lagrangian  which can be promoted to a four dimensional one
\begin{equation}
\mathcal{L}'_4=-\frac R2 +\partial_\mu U\partial^\mu U+\frac{e^{-4U}}{4}\,\omega_\mu\omega^\mu+\frac{e^{-2U}}{2}\,\partial_\mu\mathcal{Z}^T\mathcal{M} \partial^\mu\mathcal{Z}-\frac{e^{-2U}}{2\lambda}\,{\rm Tr}(\mathcal{M})+\frac{e^{-2U}}{\lambda}\,,
\end{equation}
Integrating out $g,\,\theta$ we find:
\begin{equation}
\mathcal{L}_4=\partial_\mu U\partial^\mu U+\frac{e^{-4U}}{4}\,\omega_\mu\omega^\mu+{e^{-2U}}\,\mathcal{L}_{n.l.}\,,
\end{equation}
where
\begin{equation}
\mathcal{L}_{n.l.}\equiv \frac{1}{\lambda}\,\left(1-\sqrt{1-\lambda\,(\partial_\mu\zeta\partial^\mu\zeta+
\partial_\mu\tilde{\zeta}\partial^\mu\tilde{\zeta})+\lambda^2\,(\partial_\mu\zeta\partial^\mu\zeta \partial_\nu\tilde{\zeta}\partial^\nu\tilde{\zeta}-(\partial_\mu\zeta\partial^\mu\tilde{\zeta})^2)}\right)\,.
\label{bilike}
\end{equation}
Notice that we still have the Heisenberg algebra of isometries.
For the case of rigid supersymmetry we find:
\begin{equation}
\mathcal{L}_4=\mathcal{L}_{n.l.}\,.
\end{equation}
The coupling of the non-linear hypermultiplet to gravity is thus described by the following Lagrangian:
\begin{equation}
\hat{e}^{-1}\mathcal{L}_4=-\frac{\hat{R}}{2}+\partial_\mu U\partial^\mu U+\frac{e^{-4U}}{4}\,\omega_\mu\omega^\mu+e^{-2U}\,\mathcal{L}_{n.l.}\,,
\end{equation}
which expands, for small $\lambda$, as follows
\begin{equation}
\hat{e}^{-1}\mathcal{L}_4=-\frac{\hat{R}}{2}+\partial_\mu U\partial^\mu U+\frac{e^{-4U}}{4}\,\omega_\mu\omega^\mu+\frac{{e^{-2U}}}{2}\,(\partial_\mu\zeta\partial^\mu\zeta+
\partial_\mu\tilde{\zeta}\partial^\mu\tilde{\zeta})+O(\lambda)\,.
\end{equation}
\section{c-maps and their duals in the rigid theory}\label{rigid}
\subsection{Tensor + scalar theory in BI form}
Let us reconsider the general form of the 2-derivative Lagrangian in 4D admitting a dual BI form \cite{Andrianopoli:2014mia}, for a single field. The Lagrangian has the general form:
 \begin{eqnarray}
\mathcal{L}' &=&  \frac{  g}{2\lambda}\left( \Lambda +\Sigma^2-\frac\lambda 2\,X\right) +{ \theta} \left(\frac14 Y -\frac \Sigma \lambda\right) +
\frac 1\lambda \left(1- \sqrt{1+ \Lambda }\right)\label{l'}
\end{eqnarray}
 where
 \begin{eqnarray}
X&\equiv& F_{ \mu \nu}F^{ \mu \nu}\\
Y&\equiv&\frac 12 F_{ \mu \nu}F_{ \rho \sigma}\epsilon^{ \mu \nu \rho \sigma}
\end{eqnarray}
and variation with respect to ${  g}$ and ${ \theta}$ ``dualizes'' (\ref{l'}) into the BI Lagrangian:
\begin{eqnarray*}
\mathcal{L}'|_{{  g},{ \theta}}& =&\frac 1\lambda \left(1- \sqrt{1+ \frac\lambda 2 X -\frac{\lambda^2}{16} Y^2}\right)={\mathcal L}_{BI}\label{bi}\,.
\end{eqnarray*}
As discussed above, we can again consider the dimensional reduction from 4 to 3 dimensions of the gauge field strength (in the case $\partial_3 A_{ \mu}=0$).
When decomposing $  \mu \to \hat\mu,3$, the kinetic and topological terms of  (\ref{l'}) reduce respectively to:
\begin{eqnarray}
X&\to& F_{{\hat\mu}{\hat\nu}}F^{{\hat\mu}{\hat\nu}}-2\partial_{\hat\mu}\zeta \partial^{\hat\mu} \zeta \label{dec1}\\
Y&\to& -2 F_{{\hat\mu}{\hat\nu}}\partial_{\hat\rho} \zeta\epsilon^{{\hat\mu}{\hat\nu}{\hat\rho}}\label{dec2}
\end{eqnarray}
However, the same terms (\ref{dec1}), (\ref{dec2}) would be obtained in the dimensional reduction of the four dimensional Lagrangian for a real scalar $\zeta$ plus an antisymmetric tensor field $H_{ \mu \nu \rho}=3\partial_{[ \mu}B_{ \nu \rho]}$ (in the case $\partial_3B_{ \nu \rho }= \partial_3 \zeta=0$), where:
\begin{eqnarray}
X&\equiv& -\frac 13 H_{ \mu \nu \rho}H^{ \mu \nu \rho} - 2 \partial_{ \mu}\zeta\partial^{ \mu}\zeta \label{tensx}\\
Y&\equiv&\frac 23 H_{ \mu \nu \rho}\partial_{ \sigma}\zeta\epsilon^{ \mu \nu \rho \sigma}\label{tensy}
\end{eqnarray}
if we identify $B_{{\hat\mu} 3}= A_{\hat\mu}$, $H_{{\hat\mu}{\hat\nu} 3}= F_{{\hat\mu}{\hat\nu}}=\partial_{\hat\mu}B_{\hat\nu 3}-\partial_{\hat\nu}B_{\hat\mu 3}$.\footnote{Note that, under the hypothesis $\partial_3B_{ { \nu} { \rho} }=0$, the 3D non-dynamical term $H_{{\hat\mu}{\hat\nu}{\hat\rho}}= \epsilon_{{\hat\mu}{\hat\nu}{\hat\rho}}\partial_3 \phi$,  vanishes for any  $\phi$. }
In this case the non-linear form of the Lagrangian is (as in \cite{Ferrara:2015ffa}):
\begin{eqnarray}
\mathcal{L}_{\mathrm{scal.-tensor}}& =&\frac 1\lambda \left(1- \sqrt{1-\lambda \left(\frac 16 H_{ \mu \nu \rho}H^{ \mu \nu \rho}+\partial_{ \mu}\zeta\partial^{ \mu}\zeta\right) -\frac{\lambda^2}{36} (H_{ \mu \nu \rho}\partial_{ \sigma}\zeta\epsilon^{ \mu \nu \rho \sigma})^2}\right)\label{bitens}\,.
\end{eqnarray}
and it can be generalized to the case of n fields on the same lines as in \cite{Andrianopoli:2014mia}.\par
We can further dualize the scalar $\zeta$ to an antisymmetric tensor. The resulting model describes two antisymmetic tensors and reads \cite{Ferrara:2015ffa}:
\begin{align}
\mathcal{L}_{\mathrm{n.lin-tensor}}& =&\frac 1\lambda \left(1- \sqrt{1-\lambda ( H_{1}\cdot H_{1}+H_{2}\cdot H_{2}) -{\lambda^2} ((H_{1}\cdot H_{2})^2-H_{1}\cdot H_{1} H_{2}\cdot H_{2})}\right)\label{tenstens}\,,
\end{align}
where we have used the convention that
$H_{i}\cdot H_{j}\equiv \frac{1}{3!}H_{i\,\mu\nu\rho}\,H_j^{\mu\nu\rho}$, $i=1,2$
and $H_{1\,\mu\nu\rho},\,H_{2\,\mu\nu\rho}$ are the field strengths corresponding to the two antisymmetric tensors.\par
Equations (\ref{bilike}) and (\ref{tenstens}) will be generalized to $2n$ scalars and $2n $ antisymmetric tensors, respectively, in the following.\par
Before proceeding with the derivation of the multi-scalar and multi-tensor non-linear actions, let us briefly recall the main facts about the relation, mentioned in the Introduction, of these descriptions to representations of the $N=2$ algebra broken to $N=1$. In \cite{Andrianopoli:2014mia} it was shown that the multi-vector field generalization of (\ref{l'}), or, equivalently, (\ref{linbi}), reproduces, for a suitable choice of the matrix $N$ in the scalar potential, the $N=2$ model of \cite{Ferrara:2014oka}. The latter features a spontaneous supersymmetry breaking to $N=1$ by virtue of FI terms, which define the matrix $N^{MN}$ in (\ref{linbi}), and which induce a constant matrix $C_A{}^B$ in the local realization of the supersymmetry algebra \cite{Ferrara:1995xi}:
\begin{equation}
\{Q_{A\alpha},\,\bar{J}_{\mu\dot{\beta}}{}^B(x)\}=2\,\sigma^\nu_{\alpha\dot{\beta}}\,T_{\mu\nu}(x)\,\delta_A^B+4\,
\sigma_{\mu\,\alpha\dot{\beta}}\,C_A{}^B\,,
\end{equation}
which is an essential ingredient in order to have spontaneous partial global supersymmetry breaking \cite{Hughes:1986dn,Antoniadis:1995vb}.  In this model the goldstino multiplet is an $N=1$ vector multiplet \cite{Bagger:1996wp}.
Other representations of the $N=2$ algebra broken to $N=1$ are possible, in which, as mentioned in the introduction, the  goldstino multiplet is an $N=1$ chiral or linear multiplet. These cases were investigated in \cite{Bagger:1994vj,Bagger:1997pi}, although only in the presence of a single chiral and tensor gauge multiplet (i.e. the goldstino one), respectively. The actions they find in the two works are (\ref{bilike}) and (\ref{bitens}), respectively. Below we generalize the actions (\ref{bilike}) and (\ref{tenstens}) to a generic number of fields. The generalization of (\ref{bitens}) is then obtained by dualizing half of the scalar fields to antisymmetric tensors.

\subsection{The multi-scalar Born-Infeld Lagrangian}\label{multifield}
In the spirit of the procedure of \cite{Andrianopoli:2014mia}, outlined in section \ref{linear}, the problem of determining the ${\rm U}(n)$-duality invariant multi-scalar BI action is that of minimizing the linearized Lagrangian density
\begin{equation}
\mathcal{L}_{\mathrm{lin.-scalar}}=\frac{1}{2}\partial_\mu \mathcal{Z}^T\,\mathcal{M}\,\partial^\mu \mathcal{Z}-\frac{1}{2\lambda}\,{\rm Tr}(\mathcal{M})+\frac{n}{\lambda}=-\frac{1}{2}\,{\rm Tr}(\mathcal{P}\mathcal{M})+\frac{n}{\lambda}\,\label{LLin}
\end{equation}
with respect to the non-dynamical scalars $g_{\Lambda\Sigma},\theta_{\Lambda\Sigma}$ contained in the matrix $\mathcal{M}$ introduced in (\ref{qual}), where we have defined the $2n\times 2n$ matrix $\mathcal{P}^{MN}$ as follows:
\begin{equation}
\mathcal{P}^{MN}\equiv \frac{1}{\lambda}\,\delta^{MN}-\partial_\mu \mathcal{Z}^M\partial^\mu \mathcal{Z}^N=
\frac{1}{\lambda}\,\left(\begin{matrix}{\bf 1}_n-\lambda\,\partial \zeta\cdot \partial \zeta^T & -\lambda\,\partial \zeta\cdot \partial \tilde{\zeta}^T\cr -\lambda\,\partial \tilde{\zeta}\cdot \partial {\zeta}^T & {\bf 1}_n-\lambda\,\partial \tilde{\zeta}\cdot \partial\tilde{\zeta}^T\end{matrix}\right)\,,
\end{equation}
and we have used the short-hand notation $\partial \phi\cdot \partial\xi \equiv \partial_\mu \phi \partial^\mu\xi$. The above tensor is manifestly covariant under the ${\rm U}(n)$ subgroup of ${\rm Sp}(2n,\mathbb{R})$.\par
We shall determine the BI Lagrangian by minimizing (\ref{LLin}) with respect to $\mathcal{M}$.
The resulting Lagrangian is
\begin{equation}
\mathcal{L}_{\mathrm{n.lin.-scalar}}=-\frac{1}{2}\,{\rm Tr}\left(\sqrt{-\mathcal{P}\mathbb{C}\mathcal{P}\mathbb{C}}\right)+\frac{n}{\lambda}\,,\label{minim}
\end{equation}
and is manifestly invariant with respect to ${\rm U}(n)$. The square root in (\ref{minim}) is defined, in the basis in which the argument is diagonal with eigenvalues $\lambda_i$, as the non-negative diagonal matrix with diagonal entries $\sqrt{|\lambda_i|}$. In our case the matrix $-\mathcal{P}\mathbb{C}\mathcal{P}\mathbb{C}$, being $\lambda$ small, is positive definite.\par
The above formula will  be derived  in two ways: Solving a constrained variational problem and using purely algebraic procedures based on matrix theory.
\paragraph{Variational derivation.}
We try to retrieve the result (\ref{minim}) using the Lagrangian  method of minimization of a function in the presence of constraints among the variables.

In our case the variables are the matrix elements of $\mathcal{M}$ and the constraints it obeys are the propriety to be a symmetric and symplectic matrix, namely
\begin{eqnarray}\label{symm}
 \varphi_2&\equiv& \mathcal{M}^T-\mathcal{M}=0\,,\\
  \varphi_1&\equiv &\mathcal{M}^T\,\mathbb{C}\mathcal{M}-\mathbb{C}=0\,.\label{symp}
\end{eqnarray}
The above mentioned method amounts to minimizing a linear combination of the Lagrangian (\ref{LLin}) together with the two constraints
$ \varphi_1$ and $ \varphi_2$, namely
\begin{equation}\label{comb}
   \frac{\partial}{\partial \mathcal{M}_{RS}}\left[\mathcal{L}_{\mathrm{lin.-scalar}}+\mathrm{Tr}\left(\frac14\lambda_1\varphi_1+\frac14\lambda_2\varphi_2\right)\right]=0
\end{equation}
where $\mathcal{L}_{\mathrm{lin.-scalar}}$ is
$$
\mathcal{L}_{\mathrm{lin.-scalar}}=-\frac12\,\mathrm{Tr}(\mathcal{P}\,\mathcal{M})+\mathrm{const.}
$$ while $ \lambda_1$ and $ \lambda_2$ are   two Lagrangian multipliers implementing the constraints (\ref{symm}),(\ref{symp}), which are antisymmetric matrices since so are the left-hand-side of equations (\ref{symm}) and (\ref{symp}).
We obtain from (\ref{comb}) in matrix notation:
\begin{equation}\label{expl}
   -\mathcal{P}+\,\mathbb{C}\,\mathcal{M}\,\lambda_1+\lambda_2=0.
\end{equation}
Let us try to solve the constrained equation setting  $\lambda_2=0$; it follows
\begin{equation}\label{emme}
   \mathcal{M}=-\mathbb{C}\,\mathcal{P}\lambda_1^{-1}.
\end{equation}
In order to find the explicit expression of $ \mathcal{M}$, we have still to compute $\lambda_1$. This is done setting together the above result with the two constraint equations (\ref{symm}) and  (\ref{symp}). Equation (\ref{symm}) inserted in (\ref{emme}) gives
\begin{equation}\label{firstsub}
  \mathbb{C}\,\mathcal{P}\lambda_1^{-1}=\lambda_1^{-1}\,\mathcal{P}\mathbb{C}
\end{equation}
while from equation (\ref{symp}) we find
\begin{equation}\label{secondsub}
  \mathcal{P}\mathbb{C}\mathcal{P}=  -\lambda_1\,\mathbb{C}\,\lambda_1,
\end{equation}
that is
\begin{equation}\label{tow}
  \left(\mathbb{C}\,\lambda_1\right)^2=-\,\mathbb{C}\mathcal{P}\,\mathbb{C}\mathcal{P}.
\end{equation}
Thus we have found the value of $\lambda_1$
\begin{equation}\label{fou}
  \lambda_1= \pm\mathbb{C}^{-1}\left(-\mathbb{C}\mathcal{P}\,\mathbb{C}\mathcal{P}\right)^{\frac12}\,.
\end{equation}
Finally inserting (\ref{fou}) in (\ref{emme}), we further retrieve the value of $ \mathcal{M}$ \footnote{The sign is chosen such that $\mathcal{M}$ be positive definite.}
\begin{eqnarray}\label{final}
  \mathcal{M}= -\mathbb{C}\mathcal{P}\,\left(-\mathbb{C}\mathcal{P}\,\mathbb{C}\mathcal{P}\right)^{-\frac{1}{2}}\,\mathbb{C}=
 -\left(-\mathbb{C}\mathcal{P}\,\mathbb{C}\mathcal{P}\right)^{-\frac{1}{2}}\,\mathbb{C}\mathcal{P}\,\mathbb{C}.
\end{eqnarray}
\paragraph{Algebraic derivation. }
In order to prove Eq. (\ref{minim}) we first determine a lower bound $\mathcal{L}_{min}$ for $\mathcal{L}_{\mathrm{lin.-scalar}}$ and then determine a symmetric symplectic matrix $\mathcal{M}_{min}$ such that:
\begin{equation}
\mathcal{L}_{\mathrm{lin.-scalar}}[\mathcal{M}_{min}]=\mathcal{L}_{min}\,.
\end{equation}
It is useful to write the Lagrangian density in the following form:
\begin{equation}
\mathcal{L}_{\mathrm{lin.-scalar}}=-\frac{1}{2}\,{\rm Tr}(\mathcal{P}\mathcal{M})+\frac{n}{\lambda}=\frac{1}{2}\,{\rm Tr}(\mathcal{P}\mathbb{C}\mathcal{M}^{-1}\mathbb{C})+\frac{n}{\lambda}=\frac{1}{2}\,{\rm Tr}(AB)+\frac{n}{\lambda}\,,
\end{equation}
where we have used the symplectic property of $\mathcal{M}$, $\mathbb{C}\mathcal{M}=\mathcal{M}^{-1}\mathbb{C}$, and have defined the following matrices:
\begin{equation}
A=-i\,\mathcal{P}\mathbb{C}\,\,,\,\,\,\,B=i\,\mathcal{M}^{-1}\mathbb{C}\,.
\end{equation}
Both matrices $A$ and $B$ are diagonalizable with real eigenvalues  and moreover $B$ squares to one:
\begin{equation}
B^2={\bf 1}_{2n}\,\,\Rightarrow\,\,\,\,\,|\lambda_i(B)|=1\,,
\end{equation}
$\lambda_i(B)$ denoting the eigenvalues of $B$.
If we denote by $B_D$ the diagonalized $B$ and $\tilde{A}$ the form of $A$ in the basis in which $B$ is diagonal,  we can write the following inequalities:
\begin{eqnarray}
|{\rm Tr}(AB)|=|{\rm Tr}(\tilde{A}B_D)|=|\sum_i\,\lambda_i(B) \tilde{A}_{ii}|\le \sum_i\,| \tilde{A}_{ii}|\le \sum_i\,| \lambda_i(A)|\,.
\end{eqnarray}
The latter sum can also be written as follows:
\begin{equation}
\sum_i\,| \lambda_i(A)|={\rm Tr}\left(\sqrt{A^2}\right)={\rm Tr}\left(\sqrt{-\mathcal{P}\mathbb{C}\mathcal{P}\mathbb{C}}\right)\,.
\end{equation}
We therefore find that (as above, the final sign assignment is chosen such that $\mathcal{M}$ be positive definite):
\begin{equation}
|{\rm Tr}(AB)|\le {\rm Tr}\left(\sqrt{-\mathcal{P}\mathbb{C}\mathcal{P}\mathbb{C}}\right)\,\Rightarrow\,\,\,
{\rm Tr}(AB)= {\rm Tr}(\mathcal{P}\mathbb{C}\mathcal{M}^{-1}\mathbb{C})\ge -{\rm Tr}\left(\sqrt{-\mathcal{P}\mathbb{C}\mathcal{P}\mathbb{C}}\right)\,.
\end{equation}
This allows us to write a lower bound for the Lagrangian:
\begin{equation}
\mathcal{L}_{min}=-\frac{1}{2}\,{\rm Tr}\left(\sqrt{-\mathcal{P}\mathbb{C}\mathcal{P}\mathbb{C}}\right)+\frac{n}{\lambda}\,.
\end{equation}
The value $\mathcal{M}_{min}$  for $\mathcal{M}$ at which the Lagrangian equals this lower bound is given by
\begin{equation}\label{minim2}
\mathcal{M}_{min}^{-1}\mathbb{C}=-\sqrt{-\mathcal{P}\mathbb{C}\mathcal{P}\mathbb{C}}\,(\mathcal{P}\mathbb{C})^{-1}\,\Rightarrow\,\,\,
\mathcal{M}_{min}=
\left(-\mathbb{C}\mathcal{P}\mathbb{C}\mathcal{P}\right)^{-\frac{1}{2}}\,\mathbb{C}^{-1}\mathcal{P}\mathbb{C}>0\,.
\end{equation}
Thus the BI Lagrangian reads:
\begin{equation}
\mathcal{L}_{\mathrm{n.lin.-scalar}}=-\frac{1}{2}\,{\rm Tr}\left(\sqrt{-\mathcal{P}\mathbb{C}\mathcal{P}\mathbb{C}}\right)+\frac{n}{\lambda}\,,
\end{equation}
and is manifestly ${\rm U}(n)$-invariant. We can write its explicit form by expanding the argument of the square root at lowest order in $\lambda$ (recall that $\lambda \ll 1$):
\begin{equation}
-\mathcal{P}\mathbb{C}\mathcal{P}\mathbb{C}=\frac{1}{\lambda^2}\left[{\bf 1}_{2n}-\lambda\,\left(\partial \mathcal{Z}\cdot \partial\mathcal{Z}^T-\mathbb{C}\partial \mathcal{Z}\cdot \partial\mathcal{Z}^T\mathbb{C}\right)-\lambda^2\left(\partial \mathcal{Z}\cdot \partial\mathcal{Z}^T\mathbb{C}\partial \mathcal{Z}\cdot \partial\mathcal{Z}^T\mathbb{C}\right)\right]\,.
\end{equation}
so that
\begin{equation}
\mathcal{L}_{\mathrm{n.lin.-scalar}}=\frac{1}{\lambda}\left(n-\frac{1}{2}\,{\rm Tr}\sqrt{{\bf 1}_{2n}-\lambda\,\left(\partial \mathcal{Z}\cdot \partial\mathcal{Z}^T-\mathbb{C}\partial \mathcal{Z}\cdot \partial\mathcal{Z}^T\mathbb{C}\right)-\lambda^2\left(\partial \mathcal{Z}\cdot \partial\mathcal{Z}^T\mathbb{C}\partial \mathcal{Z}\cdot \partial\mathcal{Z}^T\mathbb{C}\right)}\right)\,,
\label{msbi}\end{equation}
For $n=1$ the matrix $A=-i\,\mathcal{P}\mathbb{C}$ has two eigenvalues $\lambda_i(A)=\pm x$, where
\begin{equation}
x=\frac{1}{\lambda}\sqrt{1-\lambda\,(\partial_\mu\zeta\partial^\mu\zeta+
\partial_\mu\tilde{\zeta}\partial^\mu\tilde{\zeta})+\lambda^2\,(\partial_\mu\zeta\partial^\mu\zeta \partial_\nu\tilde{\zeta}\partial^\nu\tilde{\zeta}-(\partial_\mu\zeta\partial^\mu\tilde{\zeta})^2)}>0\,,
\end{equation}
and thus ${\rm Tr}\sqrt{-\mathcal{P}\mathbb{C}\mathcal{P}\mathbb{C}}= {\rm Tr}(|A|)=2x$ so that we find (\ref{bilike}).

\subsection{Coupling to gravity\\}
Just as we did in the two-scalar case, we can write the multiscalar non-linear Lagrangian coupled to gravity. It is
\begin{equation}
\hat{e}^{-1}\mathcal{L}_4=-\frac{\hat{R}}{2}+\partial_\mu U\partial^\mu U+\frac{e^{-4U}}{4}\,\omega_\mu\omega^\mu+e^{-2U}\,\mathcal{L}_{\mathrm{n.l.-scalar}}\,,
\end{equation}
where $\mathcal{L}_{\mathrm{n.lin.-scalar}}$ is given by (\ref{msbi}), with  $\eta_{\mu\nu}$ replaced by the space-time metric $g_{\mu\nu}$. This action describes the c-map of $n$-vector BI action.

\subsection{Dualizing scalars into tensors}\label{tens}
In the absence of gravity, the non-linear scalar Lagrangian (\ref{minim}) exhibits shift symmetries associated with the  $2n$ scalars $\mathcal{Z}^M$. This is also apparent in the linearized form of the Lagrangian (\ref{LLin}). This allows us to dualize all the scalars into tensor fields. To this end it is convenient to work with (\ref{LLin}) and to write:
\begin{equation}
\mathcal{L}'= \frac{1}{2}\,\eta^T_\mu\mathcal{M}\eta^\mu-\frac{1}{2\lambda}\,{\rm Tr}(\mathcal{M})+\frac{n}{\lambda}-\mathcal{H}^T_\mu(\eta^\mu-\partial^\mu\mathcal{Z})\,,
\end{equation}
where we have suppressed the symplectic indices and $\mathcal{H}_\mu\equiv(\mathcal{H}_{M\,\mu}),\,\eta_\mu\equiv(\eta^M_{\mu}) $.
Upon variation of $\mathcal{L}'$ with respect to $\mathcal{H}_\mu$ we get back (\ref{LLin}), while by varying $\mathcal{L}'$ with respect
to $\mathcal{Z}^M$ we find the condition $\partial^\mu\mathcal{H}_{M\,\mu}=0$ which implies that, locally,
\begin{equation}
\mathcal{H}_{M\,\mu}\equiv \frac{1}{3!}\epsilon_{\mu\nu\rho\sigma}\,H_M^{\nu\rho\sigma}\,, \quad\mbox{where } H_{M \mu\nu\rho}= 3 \partial_{[\mu}B_{M\,\nu\rho]}\,.
\end{equation}
The equations obtained by varying $\mathcal{L}'$ with respect to $\eta_\mu^M$ are:
\begin{equation}
\mathcal{M}\eta_\mu=\mathcal{H}_\mu\,\,\Rightarrow\,\,\,\,\,\eta_\mu=\mathcal{M}^{-1}\mathcal{H}_\mu\,.
\end{equation}
Replacing the solution to the above equation in $\mathcal{L}'$, up to total derivatives we find:
\begin{equation}
\mathcal{L}'_0= -\frac{1}{2}\,\mathcal{H}^T_\mu\mathbb{C}^T\mathcal{M}\mathbb{C}\mathcal{H}^\mu-\frac{1}{2\lambda}\,{\rm Tr}(\mathcal{M})+\frac{n}{\lambda}=-\frac{1}{2}\,{\rm Tr}(\hat{\mathcal{P}}\mathcal{M})+\frac{n}{\lambda}\,,
\end{equation}
where now the $2n\times 2n$ matrix $\hat{\mathcal{P}}^{MN}$ is defined as follows:
\begin{equation}
\hat{\mathcal{P}}^{MN}\equiv \frac{1}{\lambda}\,\delta^{MN}+(\mathbb{C}\mathcal{H}_\mu)^M (\mathbb{C}\mathcal{H}^\mu)^N\,,
\end{equation}
The non-linear theory is obtained by minimizing the action with respect to the matrix $\mathcal{M}$. This can be done along the same lines as in Sect. \ref{multifield}, thus obtaining:
\begin{equation}
\mathcal{L}_{\mathrm{n.lin.-tensor}}=-\frac{1}{2}\,{\rm Tr}\left(\sqrt{-\hat{\mathcal{P}}\mathbb{C}\hat{\mathcal{P}}\mathbb{C}}\right)+\frac{n}{\lambda}\,,
\end{equation}
which is manifestly ${\rm U}(n)$-invariant. For $n=1$ the above Lagrangian reduces to Eq. (\ref{tenstens}).

\section*{Conclusions}
 In this investigation we considered the c-map of non linear theories of vectors fields and their
c-map counterparts. In doing so multi-fields non linear scalar and tensor theories are obtained of the type considered in \cite{Kuzenko:2000uh,Ferrara:2015ffa}. The c-map can be extended by coupling these non linear theories to gravity then obtaining a deformation of Quaternionic-K\"ahler manifolds of N=2 theories.
It would be interesting to discuss the supersymmetric extensions of these theories, at least for the $N=1,2$ cases. In
order to achieve this a non linear constraint preserving the lower supersymmetry should be found.

\section*{Acknowledgements}
We are grateful to P.Aschieri, B.L. Cerchiai, M.Porrati, S.Theisen, A.Sagnotti and A.Yeranyan for discussions and collaborations on related subjects. The work of S.F.  was  supported in part by INFN-CSN4(I.S.GSS).

\end{document}